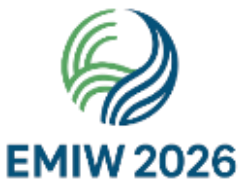

# Toward reliable and high-resolution resistivity imaging with Controlled Source EM in land – densifying arrays as a key for success - 10 years of progress for geothermal and mineral resources exploration

F. Bretaudeau[1], S. Védrine[1], C. Patzer[2], B. Kim[1], F. Dubois[1], U. Autio[2], J. Kamm[2], M. Darnet[1]
[1]BRGM (French Geological Survey), f.bretaudeau@brgm.fr
[2]GTK (Geological Survey of Finland) cedric.patzer@gtk.fi

## Summary

The Controlled Source Electromagnetic (CSEM) method aims to image electrical resistivity at intermediate depths (0–3 km) for geothermal, mineral, and groundwater exploration. It was developed both as a deeper extension of DC resistivity (ERT) and as an active alternative to magnetotellurics (MT), capable of overcoming MT's limitations in challenging environments (noisy areas or weak natural signals). As demonstrated in marine settings, CSEM can provide high resolution and complementary illumination compared to MT.
However, due to the significant logistical effort required for land deployment, most land-based CSEM surveys have used only a few transmitter positions (often one or two), effectively resembling CSAMT configurations. This limited source coverage results in poor resolution and unreliable resistivity models, where the acquisition footprint can be difficult to distinguish from actual geology.
In this work, we review ten years of progress in land CSEM and demonstrate that dense arrays—with dozens to hundreds of synchronized sources and receivers—are key to achieving the high-resolution imaging originally expected. Despite greater logistical demands, we show that such surveys can be conducted with reasonable effort and cost, making them accessible for both industrial and academic applications. Furthermore, CSEM can be combined with MT at minimal additional cost, improving model constraints and reducing exploration risk.

**Keywords:** CSEM, dense arrays, multi-TX, exploration, high-resolution, joint CSEM/MT

## Introduction

Controlled Source Electromagnetic (CSEM) methods for intermediate depths (~0–3 km) have been developed for as an extension of DC Electrical Resistivity Tomography (ERT) and as a complement to magnetotelluric (MT) techniques. As MT and DC-ERT, CSEM aims at producing intermediate depth resistivity models for a large range of applications in geothermal exploration and volcanology, mineral resource, gas, or underground water resource exploration.

This development was primarily driven by the need to overcome MT limitations in challenging environments, including marine settings, electrically noisy populated areas, and regions lacking sufficiently strong natural electromagnetic signals (Constable and Srnka 2007, Streich et al. 2011, Balasco et al. 2022, Védrine et al. 2023).

Early land CSEM implementations were essentially CSAMT-like configurations with a single transmitter position (relying on a far-field assumption or nearly far-field), whereas it is not the case in marine CSEM. As demonstrated by marine CSEM, the use of near-field and transition-zone CSEM with a large number of source position should offer enhanced resolution, greater investigation depth, increased sensitivity, and complementary information, as the current is not flowing only horizontally in the underground, and the number of independent data increases with the number of source-receiver pairs (Ntx * Nrx instead of only Nrx).

However, the logistical effort for land CSEM is significantly greater than for MT. Although CSEM receiver setups can be less demanding than MT, field mobilization, deployment, and maintenance of dozens to hundreds of synchronous receiving units remain challenging. Additionally, powerful current transmitter units are required, and large transmitter dipoles or loop sources must be deployed and synchronized with all receiving units to illuminate the subsurface with multiple directions. Consequently, most early published near-field CSEM studies were limited to only one or two transmitting positions (Schaller et al. 2018, Bretaudeau et al. 2021, Cai et al. 2022, Védrine et al. 2023, Morbe et al. 2026), effectively approaching CSAMT configurations. This resulted in major drawbacks: poor and unbalanced illumination, source statics, spatial aliasing, and ill-conditioned inverse problems. Regardless of the CSEM variant employed, these limitations produced resistivity models in which the measurement array footprint is difficult to distinguish from actual geological features (Rulff et al. 2025, Bretaudeau et al 2021, 2022, and Figure 2).

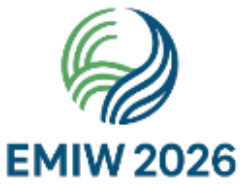

Through synthetic modeling and both ancient and recent field examples (Kim et al, 2026, Védrine et al. 2026), we demonstrate that increasing spatial coverage density to achieve an equilibrium between sources and receivers—reaching dozens to hundreds of each—was the key factor for successful land CSEM, finally delivering the resolution initially expected, comparable to what DC-ERT is providing for shallow targets (Figure 1). We further show that this approach can be implemented with realistic costs and field efforts, making it accessible for industrial applications in geothermal, mineral, and water resource exploration.

Finally, we emphasize that dense-array CSEM need not be an alternative to MT; on the contrary, MT data can be acquired jointly with CSEM for marginal additional cost, thereby increasing resistivity model constraints and resolution while reducing exploration risks (Figure 3).

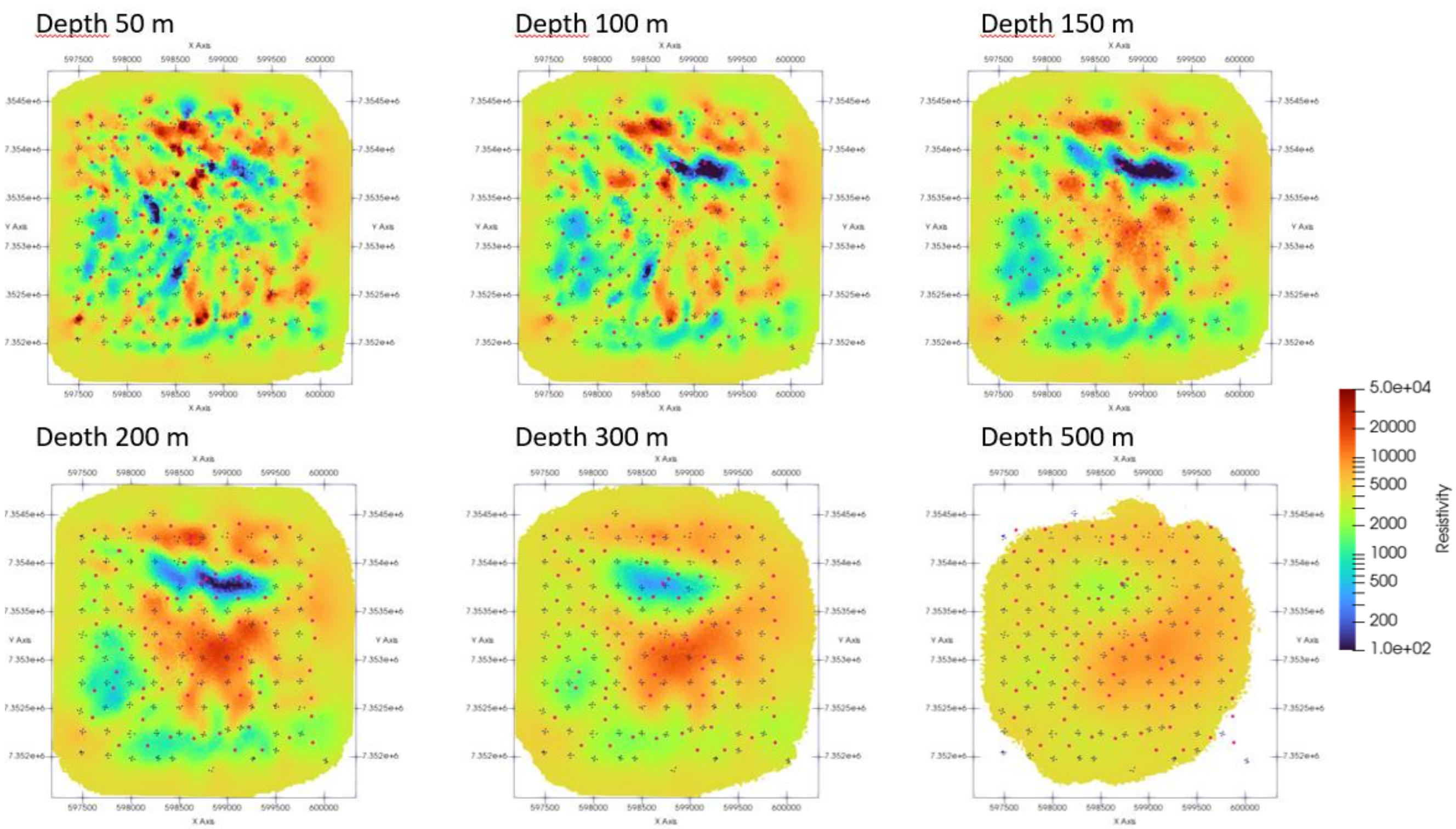

*Figure 1: 3D resistivity inversion of a CSEM data (Kuusamo dataset 2025 – Finland - UNDERCOVER project) with 100 receivers (black dots) and 92 transmitter electrodes positions (red dots)*

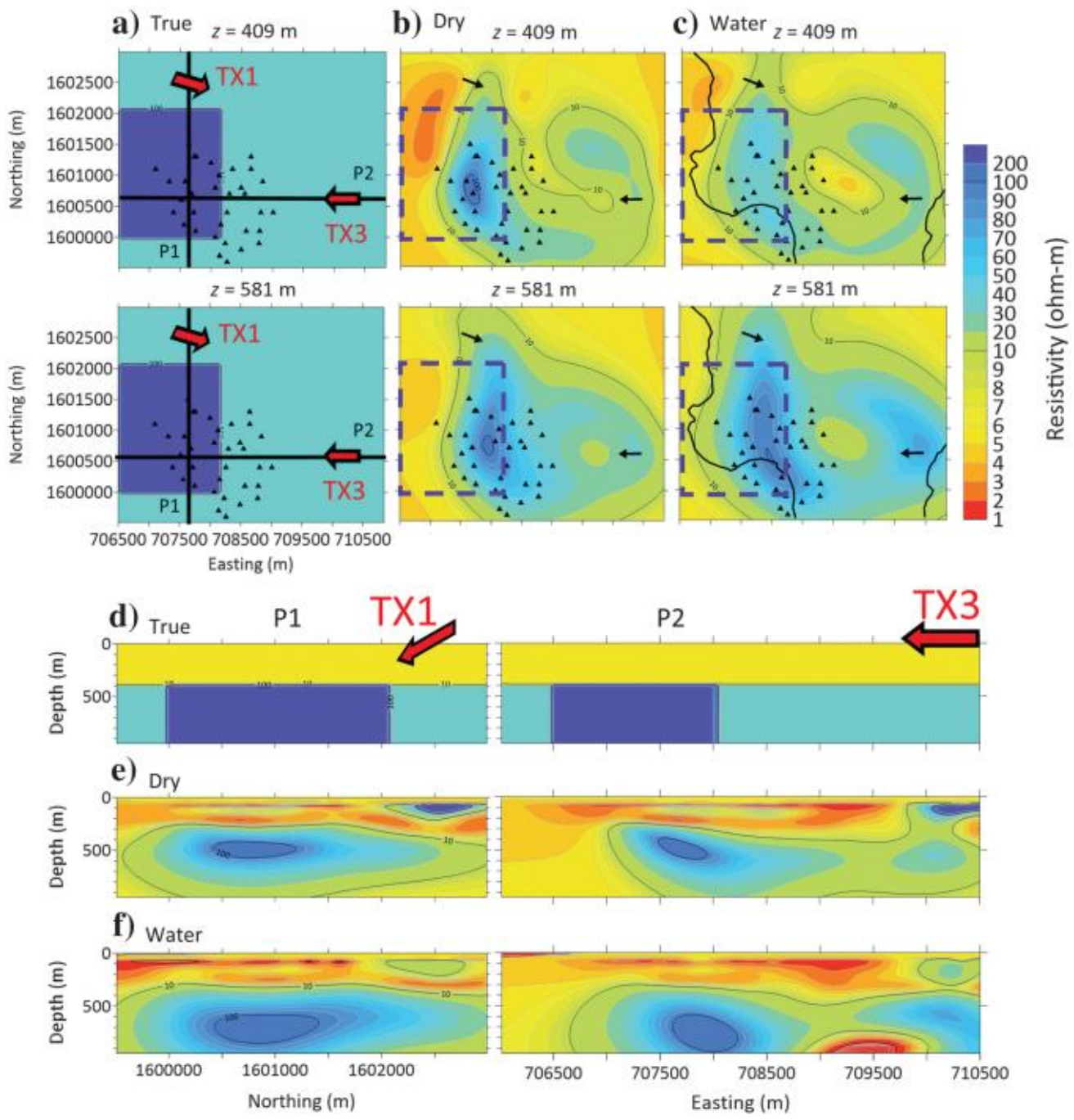

*Figure 2: Synthetic example of 3D inversion illustrating the model distortion when using only 2 transmiter (Védrine et al.2023)*

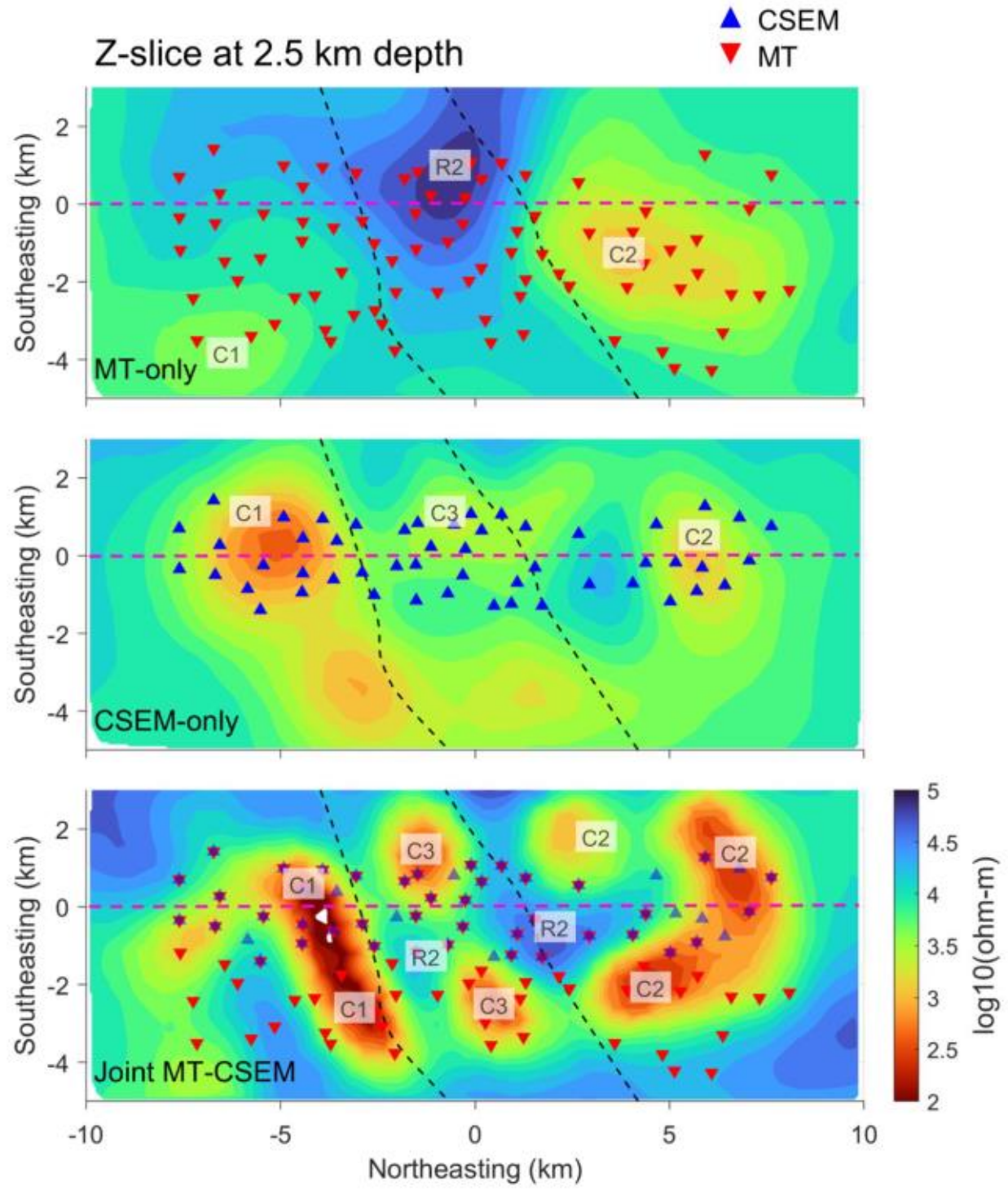

*Figure 3: Separated and joint CSEM and MT data 3D inversion (Kuusamo data set 2023) – (Védrine et al. 2026). MT provides deep information, CSEM high resolution for the first 3km, and joint inversion provide the most constrained model.*

## Acknowledgements

The material presented is coming from various former French and European research programs of those last 10 years driven by BRGM (ANR CANTARE, H2020 IMAGE, Ademe DeepEM, Ademe GEOSCAN-Paris and GEOSCAN-Arc, EU SEEMS DEEP, ANR DENHyMS) as well as studies funded by French regional and local exploration programs (Anses d'Arlet Martinique, INTERREG TEC, …) and by private companies (Lithium de France). The most recent results presented are funded by the Horizon Europe Research and Innovation Program through the project UNDERCOVER (Grant Agreement No.101177528). The authors also want to thank *Latitude 66* for their logistical and scientific support in this last one and to the numerous field crew members and contributors.